# ICFP 2020 Post-Conference Report

Stephanie Weirich, General Chair
Benjamin C. Pierce, Virtualization Chair

## 1 Overview

This document describes the ICFP 2020 virtual conference, including the planning process and the criteria that informed its design, plus feedback from the post-conference survey. It is intended to provide a record of the event and give advice to future organizers of virtual conferences.

ICFP 2020 was originally planned to take place at the Hyatt Regency Jersey City on the Hudson in Jersey City, New Jersey, August 23-28, 2020. In May 2020, the ICFP organization committee canceled the existing hotel contract and converted the conference to a virtual event, which took place online using the Clowdr platform.  The [ICFP website](#) includes the schedule of all events that took place during that week, including the Haskell Symposium and the workshops PLMW, Erlang, TyDe, HOPE, miniKanren, OCaml, ML, Scheme and HIW.

The [proceedings of ICFP 2020](#) were published in the gold open access journal PACMPL. The [front matter](#) includes information related to the selection of the ICFP program, including the number of papers submitted, review process, and acceptance rate.

The technology underpinning ICFP 2020 was a new integrated virtual conference platform called [Clowdr](#), under development since Spring of 2020. Clowdr offered easy all-in-one-place access to the conference program, the video streams and Zoom rooms for the talks, Q&A sessions, posters, and industrial sponsor meet-and-greets, plus text chats associated with the talks, other public text channels, direct text messaging, and lightweight ad hoc videoconferencing.

Some **high-level take-aways** from our experience with ICFP 2020:
- **Our approach to Q&A after talks was both highly appreciated and highly criticized***.* The most important scheduling innovation (originally forced by our very tight daily schedule) of ICFP 2020 was putting the live Q&A for each talk *in parallel* with the following talk, in a separate video chat space. The major upside of this arrangement (which many participants highly valued) was that Q&A sessions could be much more in-depth; it also significantly simplified the logistics of streaming talks, since a whole session's worth of talks could be packaged into a single chunk and uploaded to YouTube to be played automatically at a given time. The downside was that people felt bad to be forced to choose between missing the next talk or missing questions entirely. There is room for further tweaking to try to get the best of both worlds, as we discuss below.
- **A *tight, mirrored* schedule is a reasonable approach to the timezone issue.** The single issue that caused the most trouble in the design and execution of ICFP was the fact that the participants were spread all around the globe. We were fairly happy in the end with our approach to this issue, which combined fairly short days (8 hours including both technical content and social time) with a mirrored schedule that repeated most of each day's content 12 hours later.
- **Leaving completely blank spaces in the program facilitates unstructured social interaction.** A key scheduling decision was to organize the program into 90-minute or 2-hour technical sessions, each followed by an hour of structured social activities and separated by half an hour of unstructured time when nothing at all was scheduled except the "hallway track" taking place in Clowdr's ad hoc video chat rooms.
- **Clowdr worked well.** Despite being somewhat rough around the edges in its current form. (ICFP was only the second major conference to use the full platform), Clowdr served the needs of the conference. In particular, some of its innovations around enabling informal social networking and contextual text chats were appreciated by participants. (Since one of us [BCP] is also a Clowdr developer, we should note for the sake of fairness that there are a number of other integrated virtual conference platforms now available. Our point here is not to push Clowdr but rather to encourage future conferences to look beyond Zoom+Slack for their platforms!)
- **Pre-conference walkthroughs were useful.** In the two weeks before the conference, we held several events where attendees (including, importantly, student volunteers) could take a tour of the platform and get familiar with its features. This was a good idea. In hindsight, our educational materials could have been more comprehensive. For example, we assumed that attendees would be familiar with YouTube features (e.g., the fact that the ICFP sessions could be paused, caught up at a faster speed, and accessed at a later time) but comments from the survey indicated that not all attendees were aware that this was possible.
- **Attendees expect recordings to be made available promptly.** While our setup for the main conference had the side benefit of making all talks immediately publicly available, we did not have this capability for the workshops, tutorials and social events. This led to confusion and disappointment, especially for attendees outside the NY time band. While mirroring did not make sense for workshops (given their size), if we had been able to provide access to the videos more promptly, those participants could have joined asynchronously via text-based discussion.

    There were actually two separate reasons we were not able to post videos promptly: (1) distributed organization (the fact that it was workshop organizers that were responsible for making and releasing recordings, rather than the ICFP A/V team) and (2) privacy concerns (we had obtained release forms in advance for releasing pre-recorded talks publicly, but had not done so for the live workshop talks and other events). Both concerns could perhaps be addressed with suitable technology: (1) automatically record

all conference, workshop, and tutorial presentations and (2) make the recordings available immediately, but only to registered and logged-in conference attendees.
- **Decent captioning took a lot of work, but was appreciated by many.** The logistics of getting good-quality captions for all the pre-recorded talks turned out to be a bit daunting, but the benefit to a significant subset of participants was also large.
- **For all their shortcomings, virtual conferences are accessible to a much wider audience.** For example, in the post-conference survey, just as many people said that they could *only* attend virtually as said that they would come either way.

# 2 Schedule

## 2.1 Design

### 2.1.1 Process

The ICFP virtualization team spent a good deal of time discussing the issues and alternatives around the ICFP schedule --- i.e. the times during each day that various parts of ICFP would occur.

We broke the scheduling task into two separate parts: the main conference (Mon-Wed) and the workshops (Sun, Thu, Fri). For the latter, we generally allowed workshop organizers to set their own schedules, encouraging them to synchronize their breaks and start their day around the same time. Separating these plans allowed the ICFP main program to adopt mirroring, which only makes sense for very large meetings and is impractical for most workshops. (In hindsight, however, we should not have allowed them to set their own starting and break times, as this led to a rather jumbled schedule on workshop days and made it hard for people to find each other outside of workshop sessions. It would have been better to give them all the same template.)

In the end, we designed a main conference schedule that was **short** (no more than 8.5 hours of social and technical content per day), **mirrored** (replicated at a time convenient for Asia), and **tight** (events were scheduled contiguously during a fairly compressed block of times).

For the main conference, we scheduled keynotes in the 9AM EDT time slot --- a "golden" hour that is accessible for almost everyone around the globe. For the rest of the conference, we asked authors to pre-record 14.5 minute talks. The videos were then streamed on a schedule (for a "watch party" effect, where many people were watching at the same time and could discuss in real time on the text chat associated with the paper) on YouTube and iQIYI and were made available asynchronously afterwards. In parallel to these talks, we scheduled two rooms for live Q&As in Zoom, immediately following each talk (i.e., each talk was followed by up to 30 minutes of questions in one or the other of these rooms), based on author availability.

Of course, the conference schedule includes more than just the times that talks are streamed. It also includes the times that *synchronous, interactive social events* happen in conjunction with these talks. These events include talk Q&As, "coffee breaks", the "hallway track" of unstructured conversations, mentoring meetups, and other community-building events (e.g., a Women in PL event). To encourage attendance, the timing of these events was interspersed into the talk schedule. We needed to make sure to schedule both talks and social events so that people know *when* they should appear in informal, interactive, online meeting spaces.

The conference schedule also included internal "breaks" to let people decompress and attend to other needs. During the day, we wanted to have at least one 1-hour break so that people could eat, in addition to shorter breaks.

| New York | Monday, 8/24 | | Tuesday, 8/25 | | Wednesday, 8/26 | |
|---|---|---|---|---|---|---|
| 9:00 AM | KEYNOTE: Audrey Tang | | KEYNOTE: Evan C. | | Prog. Contest Report | |
| 10:00 AM | Coffee break in Clowdr | Info desk | Coffee break in Clowdr | Info desk | Awards/Chair reports | |
| | Social 1 | SRC posters | Social 3 | CARES | Coffee break in Clowdr | Info desk |
| 11:00 AM | Clowdr training | | Excursion: Art Gallery | | mini-session (2 talks) | |
| | session 1 | | session 3 | SRC talks | session 5 | |
| 12:00 PM | | | (JFP) | | | |
| 1:00 PM | Coffee break in Clowdr | Info desk | Coffee break in Clowdr | Info desk | Coffee break in Clowdr | Info desk |
| | Women in PL | | LGBTQ | | Virtualization | |
| 2:00 PM | | | | | feedback | |
| | session 2 | | session 4 | | session 6 | |
| 3:00 PM | | | | | | |
| 4:00 PM | | | | | | |
| | Social 2 | PL game | Social 4 | Trivia | Social 5 | Industrial |
| 5:00 PM | | | | | | reception |

Main program schedule (NY timeband)

| Beijing/Taipei | Tuesday, 8/25 | | Wednesday, 8/26 | | Thursday, 8/27 | |
|---|---|---|---|---|---|---|
| 9:00 AM | Women in CS | | CARES | | awards/chair report | Virtualization |
| | | | Mind the Title | | (recorded) | feedback |
| 10:00 AM | Clowdr training | coffee break | Olivier Danvy | | coffee break | |
| | Asia 1 | | Asia 3 | SRC talks | Asia 5 | |
| 11:00 AM | | | | | | |
| | | | | | | |
| 12:00 PM | meal break | | 1-n mentoring | | | |
| | | | Guy Steele | | 1-n mentoring | |
| 1:00 PM | 1-n mentoring | | meal break | | Matthias | Felleisen |
| | Ranjit Jhala | | | | meal break | |
| 2:00 PM | coffee break | | coffee break | | | |
| | Asia 2 | | Asia 4 | | Asia 6 | |
| 3:00 PM | | | | | | |
| | | | | | | |
| 4:00 PM | | | | | | |
| | Online lecture panel | | Some Proverbs in Type Theory | | | |
| 5:00 PM | | | | | | |

Main Program schedule (Asia band)

## 2.1.2 Considerations for planning the schedule

In planning the ICFP schedule, the virtualization team considered the following issues:
- Access - How can we allow as many people as possible to participate?
- Community - How can we make sure that people meet each other?
- Simplicity - How hard is it to understand the schedule and find the events/talks?
- Fatigue - How many hours will attendees spend online each day?
- Organizational overhead - How much are we asking from the organizers (and how many organizers do we need)?

These considerations highlighted several key trade-offs in the design:
- *Mirroring vs. Single instance*
  Because we plan to have pre-recorded talks, we can "easily" run the entire program twice -- should we?
  **Mirroring PROs**: more access, especially for people in Asia/Australia. (By default ICFP 2020 ran on a schedule convenient to NY's timezone because that was where the organizers were located. ICFP 2021 will be in the Korean timezone.)
  **Single Instance PROs**: Authors may be required to participate in more than one Q&A with a mirrored schedule, which may occur at an inconvenient time for them. Mirroring might also fragment the community. Mirroring has a higher organizational overhead: need a team from Asia to organize social events and manage the conference mirror. Both teams need to communicate with each other to plan the event (which itself is also made challenging by timezones).
  **ICFP Choice**: We chose to mirror ICFP 2020. We recruited Sukyoung Ryu (ICFP 2021 General Chair) to join our virtualization team and assist us in mirroring. Although this represented a significant additional task for her, it did give her a valuable preview of virtual conference organization for next year. Sukyoung recruited a Asia-based SV co-chair, student volunteers and AV team (for iQIYI) and put together the social program for the Asia band.

- *Short vs. long schedule*
  Should we constrain everything to ≤8 hours or spread events to 12 hours?
  **Short PROs**: Zoom fatigue is a real thing. A short schedule might compress community participation, allowing more interaction. (If people are spread too thin, they may not run into each other.) Might make it easier for attendees to feel like they have participated in "all" of ICFP.
  **Long PROs**: If we *didn't* have mirroring, spreading the schedule out would provide more access. Talks could be longer. More time for social events between talks/ technical sessions. Maybe more flexibility for people to attend to family responsibilities, pressing work stuff, etc.
  **ICFP Choice**: We used a shorter schedule, mostly to foster interaction. This led to a quite compressed talk slot time of just 15 minutes.

- *Tight schedule vs. distributed time blocks*
  Should we schedule the program within a 12 hour day or consider something like ICSE's 24 hour schedule of three 3-hour time blocks evenly spaced through the day?
  **Tight PROs**: Easier for people to figure out and for organizers to set up. Without mirroring, spaced time blocks would mean everyone will have parts of the conference that they cannot attend. With a well-chosen tight schedule, some (most?) people may be able to make it to the entire conference.
  **Distributed PROs**: More access, even without mirroring. No timezone is "favored".
  **ICFP Choice:** Because we had mirroring, we liked the tighter schedule. It seemed simpler for organizers and easier for participants to manage.

## 2.2 Data

In this section we present some quantitative information related to attendance, time zones, and scheduling, together with related results from the post-conference survey.

### 2.2.1 Main-program engagement

As of September 3rd (a week after the end of ICFP), there were the following viewer counts on YouTube. (All non-China based participants were directed to the YouTube site via Clowdr; participants in China could watch the stream via iQIYI.) The keynotes were livestreamed while the six sessions appeared as YouTube "Premieres", each repeated in both the NY and Asia timebands. The contents of the sessions were identical in the timebands, i.e. the same video played both times.

```
Keynote: Audrey Tang        1.6K views
Keynote: Evan Czaplicki     2.3K views
Awards and reports          1.6K views
Session 1 - 1K (NY), 224 (Asia)
Session 2 - 1K (NY), 333 (Asia)
Session 3 - 421 (NY), 116 (Asia)  - JFP session
Session 4 - 447 (NY), 190 (Asia)
Session 5 - 336 (NY), 152 (Asia)
Session 6 - 429 (NY), 140 (Asia)
```

The keynotes maxed out at 287 and 250 concurrent viewers respectively. Our most viewed session got ~570 unique views (with around ~150 concurrent). In addition, the ICFP playlist received 132 plays on the Chinese streaming site Bilibili.

To compare, videos from ICFP 2019 were uploaded to YouTube in December 2019. Although a few individual talks had more than 1k views on September 3rd, the keynote view counts were each around 150. People aren't very aware of the availability of the recordings from the most recent physical event.

### 2.2.2 Mirroring

The majority of participants in the survey reported that they either exclusively or mostly attended the NY time band, at 318 out of 402 respondents (about 80%). In contrast, only 36 (less than 10%) reported exclusively or mostly Asia. The remainder attended both bands about equally.

That said, 109 people (out of 405) reported that mirroring was extremely important to them, while only 17 preferred no mirroring. The reasons cited against mirroring were concern about splitting the audience, confusion about the schedule, a preference for asynchronous video, a preference for a specific timezone, or a belief that mirroring was unnecessary for recorded sessions. In hindsight, we could have done a better job communicating that the YouTube streams would be accessible after the premiere and that the Q&A sessions would not be recorded.

We also asked attendees about the tradeoff between live talks and mirroring. The overwhelming majority preferred mirroring, with only a small minority (26 out of 407) indicating that live talks were so important to them that they would give up mirroring for them. A large number (173 out of 407, 43%) preferred pre-recorded talks to live talks anyway.

Overall, the majority of participants somewhat or strongly agreed that the quality of online video online video presentations at ICFP was at least as high as the quality of physical conference talks (236/388, 60%). The number that somewhat or strongly disagreed was under 20%.

### 2.2.3 Q&A breakout rooms

For the Q&A sessions, 613 unique Zoom names attended at least once. This produced a total of 39,320 minutes spent in Q&A overall, for an average of 64 minutes per registrant. The median length of time spent per person, per session was 13 minutes. A few people camped out in the Q&A room and stayed for multiple Q&A sessions. Of the total amount, 6039 minutes (15% of total) were during the Asia band.

Excluding the special session devoted to presentations of papers recently accepted for the *Journal of Functional Programming,* we had 36 Q&A sessions in the NY time band and 24 in the Asia time band; the discrepancy in numbers is because some authors only wanted to have a single Q&A in their "home" timeband. (All of the JFP authors participated in the NY band and none in the Asia band.)

In the post-conference survey, our question about the Q&A room experience overall received 389 responses. Of these, 37% percent said they were very happy with ICFP's arrangement ("Future virtual conferences should do it this way"), 24% were ok with it, and 10% would have preferred something else.

Comments in favor:
- Liked the depth of discussion afforded by the extended Q&A
- Found the Q&A fun and useful
- Liked that the author could join the text chat during the talk
- "This was one of my favorite things about this format"
- Appreciated that missed talks could be watched doublespeed. "This kind of effective time use surrounding the Q&A is not possible in a physical conference."
- "I enjoyed the Q&A sessions I attended more than the ones at in-person conferences in the past"
- Liked that pipelining the schedule made more time for synchronous social time
- "I *REALLY* loved the Q&A format. It was more like a fireside chat, and it was so great to interact in that setting.
- Liked that could hear from multiple co-authors, not just presenter
- It felt a lot like the informal knots of people who gather around the speaker after a session at a physical conference, except that many more people got to listen in.

Against:
- Most common complaint: Disliked having to miss the next talk (or having to choose)
- Expressed a preference for short Q&A in the livestream before going to a separate room
- Got too caught up in the discussion (more than 30 minutes) and missed a talk they wanted to see
- Missed seeing others' questions for talks that they have only a passing interest in
- Wanted more social structures for asking questions in the Q&A
- Had trouble with Zoom audio
- Would prefer all talks in the session to have one joint Q&A after the session concludes
- Thought questions were less moderated if they were not in the main stream
- Found it difficult to find the right Q&A room, even with SV support
- Didn't go to them because they overlapped the next talk
- Thought it took too much overhead time to switch to the Q&A session (wanted them to start immediately when the talk ends, not waiting for the "scheduled" time)
- Authors may not want to discuss for such a long time (30 minutes)
- Didn't like that there were many more Q&As in NY band than Asia band
- Thought text channels were just fine for discussion and disliked video-based Q&A
- Lost the "community-wide aspect" of the Q&A

Many comments suggested logistical changes:
- Wanted access to the prerecorded talks beforehand so they could watch them before the Q&A sessions (multiple respondents)
- Wanted to allow Q&A sessions to continue for more than 30 minutes
- Suggested a lightning talk session to help people plan their time
- Wanted the default "up-next" to be the Q&A session for the talk instead of the next paper (which would start in a different room)

- Text-only Q&A so that one could pay attention to it and the next talk simultaneously
- Wanted to be able to ask questions in advance
- Wanted the Q&As to be recorded
- Wanted a 10 minute public Q&A followed by a 20 minute overlapping Q&A
- Wanted longer slots (20 minutes) but shorter talks (10-13 mins)
- Wanted contact info for everyone in the chat for follow up
- Wanted to put multiple Q&As in parallel so that they don't overlap the talks
- Wanted to overlap Q&As with social time
- Wanted a way to watch the Q&A on YouTube instead of Zoom
- Wanted moderators to produce a summary of the Q&A session, posted in the text channel
- Wanted sli.do (anonymous voting on question order in Q&A)
- 2-3 minute pre-recorded Q&A in the main stream
- Wanted extended discussions only in an ad hoc manner

### 2.2.4 Talk length

The post conference survey asked about the ideal length of talks. The most popular answer was the category that included the length of ICFP talks (14.5 minutes), but a significant number wanted longer talks.

| # | Answer | Count | % |
| --- | --- | --- | --- |
| 1 | < 10 minutes | 4 | 1.05% |
| 2 | 10-13 minutes | 16 | 4.19% |
| 3 | 14-17 minutes | 193 | 50.52% |
| 4 | 18-20 minutes | 129 | 33.77% |
| 5 | > 20 minutes | 40 | 10.47% |
|   | Total | 382 | 100% |

## 2.3 Discussion

### 2.3.1 Pre-recorded stream plus Q&A breakout rooms

Having the main program sessions pre-recorded both helped and hindered the organizational progress. On one hand, it made mirroring simpler --- the YouTube sessions for both time bands were set up by a single AV coordinator. Furthermore, a significant Zoom outage on the first day of the conference did not affect the main talk stream. On the other hand, the timing of the YouTube Premiere feature was not precise and directing people to the appropriate Q&A rooms (when available) was only possible with dedicated Clowdr support.

Overall, if we did it again, we would definitely release the talks in advance in addition to setting up the timed stream (which is important for the "watch party effect"). We would also strongly consider adding 3-5 minutes between talks. This would give a bit of breathing room to the schedule. It would also give people a chance to decide if they want to continue with the Q&A or watch the next talk. (There are some logistical concerns with this plan, but SPLASH 2020 will be trying a variant of it. We will be very interested to see whether it works well!)

Having all the Q&As for a session in parallel with each other at the end of the session is also an intriguing idea, but would cut into social time. It would also prevent attendees from joining multiple Q&As.

### 2.3.2 Tight schedule

From an organizer point of view, compressing the schedule helped combat fatigue during the conference and gave the team a chance to "reset" between bands. A full-on 24 hour schedule would have been difficult and would have required a 50% larger organization team, which would have been a large "ask" for a relatively small community like ICFP. Larger conferences may make a different calculation.

# 3 Fostering Social Interaction

Because ICFP was a virtual event, a major concern was that there would be fewer opportunities for social interactions. In this section we explain and evaluate our approach to this concern.

## 3.1 Design

### 3.1.1 Organized events

The organized social events provided a structured way for participants to interact. These events ranged from workshops & meetups for specific community groups (women, LGBTQ), mentoring activities (1-n, mind the title), fun experiences (card & trivial game), tutorials, and ShutdownPL a half-day event focused on increasing the participation of minorities in PL research.

The Asia band included more structured events than the NY band, because we were worried that there would not be enough participants there to achieve critical mass for unstructured events.  This was a good call, in retrospect.

### 3.1.2 Ad hoc chat rooms

The Clowdr platform also provided attendees with the possibility to join ad hoc chat rooms. The conference organizers encouraged this behaviour by dedicating space for this activity in the conference program, running training sessions for the platform, specifically asking ICFP community members to make time for these events, and reminding attendees to join the chat rooms at the ends of the technical sessions.

Overall, the usage of the Clowdr video chat rooms was higher than expected (though still lower than the amount of unstructured interaction we are used to from physical conferences). During the course of the week, 342 different people spent at least 1 minute in the Clowdr video chat rooms, almost 100 people spent two hours or more, and more than 50 people spent four hours or more. The heaviest user spent more than 20 hours in the chat rooms during the week.

### 3.1.3 Mentoring program

Talia Ringer proposed and organized a mentoring program which matched up registrants for both short term (one meeting during the ICFP week) and long term (multiple meetings over the year) mentoring. This program proved to be immensely popular---over 360 attendees signed-up (either to be a mentor, receive mentorship or both).

## 3.2 Data

A selection of the free-form responses to "Do you have any feedback about associated events?":
- All of the events I attended were great!
- They were great, and we should keep them, even in person.
- An LGBTQ meeting was essential for me in getting involved with the social aspect of the conference
- I appreciated the diversity, even if I didn't attend many, due to burnout. I would like to see this continue.
- They were great! Please keep them in at least this diversity.
- WE WANT MORE
- It was great to see how much effort had gone into making this virtual conference as social as possible.
- I would have liked to attend more of these. It was much easier to overcome the awkward barrier of joining a video chat if there was a pre-defined (if broad) topic to the conversation. I also liked this with informal chat rooms that had more-or-less descriptive titles (instead of just "Chat Room X").
- I'd like more casual events like the trivia! Since it was impossible to go for dinner or drinks with fellow attendees, it provided a nice environment to chat, joke around and socialize.
- These social events were the highlight of the conference for me. ShutDownPL was especially fantastic. This should be a regularly-occurring event so that we can keep up the momentum around these deeply important conversations!
- I think it was good to have structured virtual interactions. At times it was a bit chaotic, but kind of worked.
- I think we had too many social events and not enough time on the main event, I hope future conferences strike a better balance. Also, people are themselves quite good at socializing, especially, when they have things to discuss, so probably there is no need to shepherd them.

- I did not attend any social events. Perhaps, I would have attended (some of them) if they were in my time zone. I took the new york midday break to move from work to home, have dinner and be ready for the afternoon session. Still I feel like virtual socializing too artificial/unreal for me.
- I felt less willing to socialise due to the remote nature of the conference. I could be working instead (particularly with time zones being as they are).
- I understand why informal chat is not recorded, to leave people able to speak freely. But it's a pity that I therefore missed some of the social sessions, either because of timezones or because of scheduling conflicts.
- I'm afraid I found the social side of this ICFP impossible to navigate.
- It was difficult to know what they all were, and the usual informal channels of asking someone were not available.
- I did not see how to join social events.

## 3.3 Discussion

Given that this was a virtual conference, we were overall pleasantly surprised by the level of participation in the social programs. Many participants got real use out of the ad hoc Clowdr chat rooms and the more structured social events had good attendance and feedback.

Overall, we think that a few key decisions led to this outcome:
- First, we made sure to leave some real breaks in the middle of the schedule where there were only unstructured events. Not only does this give people a much needed break, but it signals the importance of these unstructured activities in the first place. These weren't simply "unscheduled" times --- we specifically included events in the program for "socializing in the Clowdr chat rooms," providing a social indication of when and where unstructured interactions should occur. (They occurred at other times too, of course, but repeatedly nudging people toward the chat spaces helped.)
- In service of these scheduled but unstructured events, we asked session chairs and organizers to make clear reminders that at the end of technical sessions. This provided a natural flow of people from the presentations to the chat rooms and a natural start to new conversations.
- Finally, we did some preparation work: we specifically contacted senior and visible community members and asked them to spend time in the chat rooms. Clowdr's interface, which displays the names of all people currently chatting, meant that these members provided additional incentives for others to join the discussion.

Some participants expressed dismay after the conference, viewing it as the best that can be done in a virtual environment, while regretting that it doesn't stand up to a physical event. We disagree with this conclusion---we as a community are only just learning how to socialize virtually, from a technological, organizational, and social standpoint. We need to develop and adopt new social protocols before these events will feel natural and train people to expect and look for these events. The platform itself needs to be better at guiding participants to finding these ad hoc interactions. We also need better support from the platform for unstructured social protocols. (E.g., the platform could do better in pushing people into groups of comfortable sizes and helping communicate nonverbal social cues between participants.)

# 4 Attendance and Accessibility

The fact that ICFP was virtual meant that in some ways it was accessible to a wider range of attendees than previous events.

## 4.1 Attendance report

Overall, 1155 people registered for ICFP, either directly through RSL or indirectly through an application for PLMW.

| **ICFP 2020 registrations** | Early | Late | Total |
| --- | --- | --- | --- |
| ACM/SIG Member | 162 | 58 | 220 |
| Nonmember | 125 | 93 | 218 |
| Student member | 147 | 23 | 170 |
| Student nonmember | 118 | 68 | 186 |
| Student volunteer | 48 | 7 | 55 |
| Other comp | 90 | | 90 |
| PLMW | 216 | | 216 |
| **TOTAL** | **906** | **249** | **1155** |

Overall, a significant fraction of the ICFP 2020 attendees registered for free (about ⅓, almost all students).

| Free | 508 | PLMW, SVs, sponsors, waivers, etc. |
| --- | --- | --- |
| Non-free | 647 | 438 nonstudents / 209 students |

The numbers more than double our pre-conference estimates for attendance in these categories and are significantly higher than the corresponding numbers for in-person ICFPs.

Compared to 2019, the number of ACM members decreased, but the number of non-member non-student registrations almost tripled. Furthermore, because ICFP 2020 was free (or nearly free) for students, the number of student attendees at ICFP saw the most significant increase.

| **ICFP 2019** | Early | Late | Total |
| --- | --- | --- | --- |
| ACM/SIG | 244 | 72 | 316 |
| Nonmember | 48 | 52 | 100 |
| Student member | 77 | 15 | 92 |
| Student nonmember | 25 | 11 | 36 |
| SV | 33 | 9 | 42 |
| **Total** | **427** | **159** | **586** |

Part of the reason that the 2020 number is high compared to 2019 is that PLMW decided to accept almost all applicants this year. (There were 275 PLMW participants in 2020.). However, even only comparing the non-student registrations (438 vs. 416), ICFP 2020 represents an increase in registration numbers.

## 4.2 Geographic distribution of attendees

We had country information for a majority of registered participants. Aggregating across continents produced the following breakdown. (NOTE: Turkey, Israel and Russia are included in Europe).

|  | Nonstudent |  | Student |  | Total |  |
| --- | --- | --- | --- | --- | --- | --- |
| North America | 201 | 45% | 266 | 51% | 467 | 48% |
| Europe | 179 | 40% | 145 | 28% | 324 | 33% |
| Asia | 60 | 13% | 97 | 19% | 157 | 16% |
| South America | 3 | 1% | 12 | 2% | 15 | 2% |
| Australia/NZ | 5 | 1% | 3 | 1% | 8 | 1% |
| **TOTAL** | **448** |  | **523** |  | **971** |  |

The analytics from Zoom and the post-conference survey response provide a similar geographic distribution.

## 4.3 Comments about accessibility in the post-conference survey

The post-conference survey asked: Please indicate how strongly you agree or disagree with the following statement: "If the COVID-19 pandemic had not happened and ICFP 2020 had been held as normal, I would have attended the meeting physically. Roughly equal numbers replied that they would not have been able to attend as who said that they would have come either way.

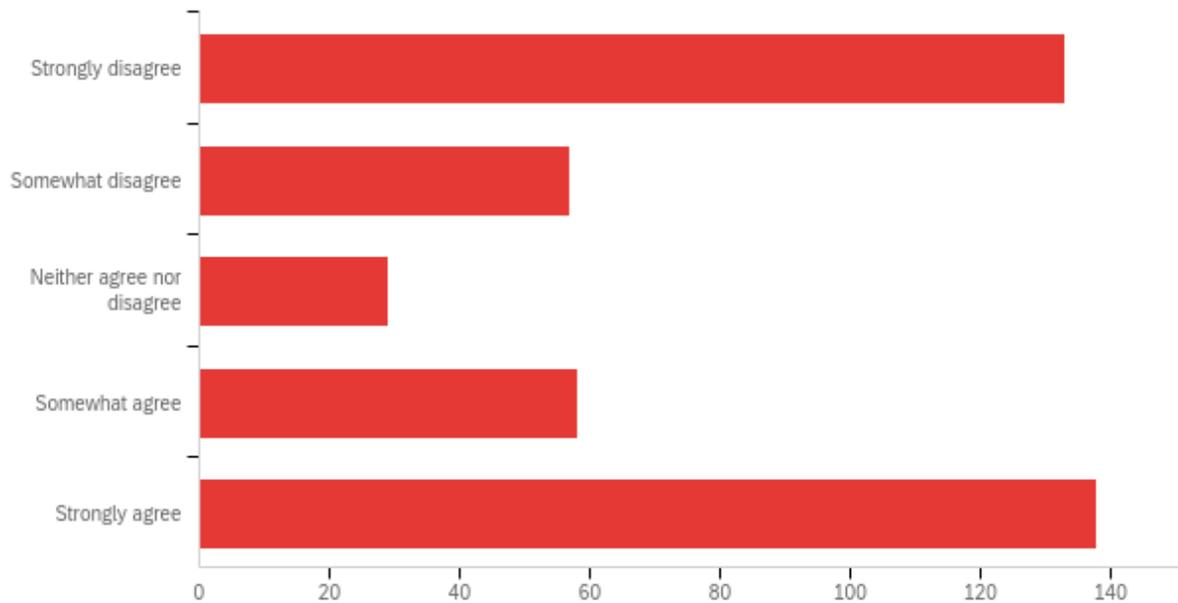

The most common reason given for not being able to attend a physical meeting was the cost of attendance in terms of money (205) and then time (159). The next two responses were carbon cost (85) and a preference to avoid travel to US/NJ (69). Other factors included difficulty in getting a visa (39), family issues (27) and health issues (26). (Respondents could select multiple answers on this question.)

## 4.4 Captioning

All talks in the ICFP main program and in the PLMW workshop were captioned by [AI Media](...), a company based in London UK that specializes in captioning services for technical events. PLMW and the ICFP keynotes used live captioning --- the live video stream was slightly delayed so that captions could be added inline. ICFP / JFP pre-recorded talks were sent to be captioned ahead of time and the caption data uploaded to YouTube before the stream began. For these, Alan Jeffrey and Lindsey Kuper went through and corrected technical terms in the caption files. After the conference was over, they made the caption files available to the authors for correction. (Our tight production schedule did not provide time for this before the event itself.)

The survey asked "To what extent did the captioning (subtitles) improve your experience of the technical talks?" The response (n=395) was positive or neutral, with a handful of respondents complaining that the captions for the live keynotes could not be disabled. (Whereas the pre-recorded talks had closed captions, which could be disabled, the live stream had open captions, which could not. This was the compromise we landed at because we were streaming to two services, YouTube and iQIYI, which made live closed captioning challenging. This technical challenge could probably be overcome with sufficient lead time and coordination with the captioning service.)

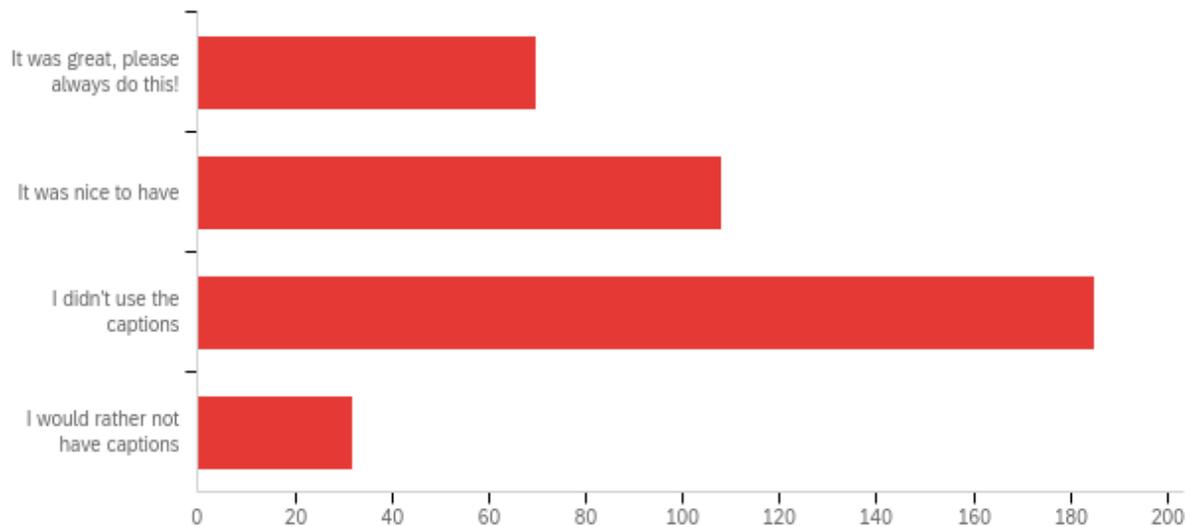

## 4.5 Registration fee

ICFP 2020 included a registration fee according to the following table. For simplicity, registration the registration fee was all-inclusive: there were no separate fees for co-located events, nor discounts for people who wanted to attend fewer than six days.

|  | On or before 8/8/20 | After 8/8/20 |
| --- | --- | --- |
| ACM or SIGPLAN Member | $75 | $100 |
| Nonmember* | $100 | $125 |
| Student member of ACM or SIGPLAN | FREE | $15 |

| Student nonmember* | $15 | $30 |

*How were these levels determined?* Registration fee levels were set in early July, a time of significant uncertainty. At this point, it was difficult to estimate both the revenues and expenses for the conference. No prior online SIGPLAN conference had charged a fee, so we did not know how these fees would affect registration numbers. Would they keep people away? Our estimates were based on numbers from ICFP 2019, a physical conference in Germany, with the understanding that they would almost certainly be wrong. Furthermore, when we set the registration fee, we had no idea how the pandemic would affect our industrial sponsors; would they still be willing or able to support a virtual event? Finally, at that point we also had not made any decisions with respect to the software technologies that we used for the virtual conference and had little understanding of the various costs of the platforms.

Instead, our registration fee was based on two pieces of guidance: the ACM task force document https://www.acm.org/virtual-conferences, which recommended that virtual conferences include a registration fee, and the results from the VR2020 and PLDI 2020 post-conference surveys, which indicated that our chosen registration levels would be in line with community expectations.

At the maximum level, our rate corresponds to less than the standard fee for a single day of a physical meeting. Given that we did not have the capability to allow per-day access to the event, it seemed reasonable to not charge more than this amount.

*Differential pricing.* What this table does not capture is that there were many ways to register for ICFP 2020 for free: Student Volunteers and PLMW attendees could register without charge and received SIGPLAN membership (if they did not have it already). We also offered a *waiver program* where anyone could request a free registration to ICFP, and all requests were accepted. Twenty-seven people received a waiver in this way. Furthermore, the Erlang Workshop used sponsorship money to set up a scholarship program to also provide free registrations for attendees.

## 4.6 Survey response to registration fee

The post-conference survey asked the following question: "Q47 - In round figures, what is the maximum registration fee that you, or your employer or institution, would have been willing to pay in order for you to attend virtual ICFP 2020, if you had known in advance what the conference would offer?"

The possible responses were:

1. Nothing: I would only have attended if there was no fee
2. 1-10 USD
3. 11-50 USD
4. 51-100 USD
5. 101-200 USD
6. 201-300 USD
7. 301-500 USD
8. I think it would be reasonable to charge the fees associated with a typical SIGPLAN physical conference
9. I don't know

There were 412 responses to this question. 90 answered that they didn't know, with 322 providing a concrete answer.[1]

Breaking the responses down by participant's role, the highest responses were at the fee levels that were set for that role. For example, the majority of students preferred $11-50 (and were charged $0-30) whereas academic researchers were split between $51-100 and $101-200 (and were charged $75-125). Professional developers were more uniformly distributed in their answers.

Notably, very few respondents selected option 1 (no fee) or option 2 ($1-10).

| # | Question | Student | Academic researcher | Industry researcher | Professional developer | Other computer i | Other | Total |
|---|---|---|---|---|---|---|---|---|
| 1 | Nothing: I would only have | | 8 | 4 | 1 | 3 | 0 | 0 | 16 |
| 2 | 1-10 USD | | 12 | 2 | 0 | 1 | 1 | 1 | 17 |
| 3 | 11-50 USD | | 68 | 16 | 0 | 13 | 0 | 5 | 102 |
| 4 | 51-100 USD | | 22 | 37 | 5 | 15 | 1 | 1 | 81 |
| 5 | 101-200 USD | | 3 | 34 | 3 | 16 | 1 | 0 | 57 |
| 6 | 201-300 USD | | 2 | 8 | 7 | 10 | 3 | 0 | 30 |
| 7 | 301-500 USD | | 1 | 4 | 1 | 1 | 0 | 0 | 7 |
| 8 | I think it would be reasonab | | 5 | 2 | 1 | 1 | 2 | 1 | 12 |
| 9 | I don't know | | 41 | 17 | 6 | 24 | 1 | 1 | 90 |

---

[1] Of course, this is a survey for people who attended ICFP 2020 and excludes anyone that stayed away because of the fee.